# Unraveling the dynamic slowdown in supercooled water: The role of dynamic disorder in jump motions

Shinji Saito[1,2,*]

[1] Institute for Molecular Science, Myodaiji, Okazaki, Aichi, 444-8585, Japan

[2] The Graduate University for Advanced Studies (SOKENDAI), Myodaiji, Okazaki, Aichi, 444-8585, Japan

*Corresponding author: E-mail: shinji@ims.ac.jp

**Abstract**

When a liquid is rapidly cooled below its melting point without inducing crystallization, its dynamics slow down significantly without noticeable structural changes. Elucidating the origin of this slowdown has been a long-standing challenge. Here, we report a theoretical investigation into the mechanism of the dynamic slowdown in supercooled water, a ubiquitous yet extraordinary substance characterized by various anomalous properties arising from local density fluctuations. Using molecular dynamics simulations, we found that the jump dynamics, which are elementary structural change processes, deviate from Poisson statistics with decreasing temperature. This deviation is attributed to slow variables competing with the jump motions, i.e., dynamic disorder. The present analysis of the dynamic disorder showed that the primary slow variable is the displacement of the fourth nearest oxygen atom of a jumping molecule, which occurs in an environment created by the fluctuations of molecules outside the first hydration shell. As the temperature decreases, the jump dynamics become slow and intermittent. These intermittent dynamics are attributed to the prolonged trapping of jumping molecules within extended and stable low-density domains. As the temperature continues to decrease, the number of slow variables increases due to the increased cooperative motions. Consequently, the jump dynamics proceed in a higher-dimensional space consisting of multiple slow variables, becoming slower and more intermittent. It is then conceivable that with further decreasing temperature, the slowing and intermittency of the jump dynamics intensify, eventually culminating in a glass transition.

# I. INTRODUCTION

Rapid cooling of a liquid below its melting point without inducing crystallization significantly slows its dynamics.[1-3] For example, the structural relaxation time increases by ~5×10$^4$-fold for a temperature decrease of ~100 K from room temperature in liquid water.[4] This slowing occurs without significant structural changes in supercooled water, in contrast to the crystallization process, in which a new peak appears in the radial distribution function (RDF), as shown in Fig. 1(a). The relationship between structure and dynamics has been extensively investigated, including studies of local geometrical orders[5,6] and localized defect modes.[7-9] Structural changes have also been studied using a cage-jump picture[10-18] associated with dynamical facilitation.[2,3,19] Machine learning techniques have recently been applied to identify structural quantities correlated with dynamics.[20-25] Nevertheless, the origin of the slowdown remains an open question.

Water plays a crucial role in a number of chemical, physical, and biological processes. It is also a unique liquid with various anomalous properties, such as significant changes in thermodynamic response functions in its supercooled state.[26-28] Extensive experimental and theoretical research has been conducted to understand these anomalies. In recent years, there has been increasing experimental[29-32] and computational[4,5,33-45] evidence in support of the liquid–liquid critical point (LLCP) scenario,[46] which attributes the anomalous properties of water to pronounced fluctuations between two liquid states derived from low-density liquid (LDL) and high-density liquid (HDL). The LLCP is considered to be located in a low-temperature, high-pressure region, e.g., $T_c$ ~ 171.5 K, $P_c$ ~ 1872 bars (Ref. [44]) based on the TIP4P/2005 model.[47] Consequently, liquid water at ambient pressure can be described by the two liquid states, hereafter referred to as LDL-like and HDL-like states associated with the LDL and HDL states, respectively. The structural characteristics of the two states have been studied previously.[4-7,26,27,38-41] The LDL-like state is characterized by tetrahedral hydrogen-bonding structures with lower local density, whereas the HDL-like state is characterized by locally distorted structures with higher local density, which facilitate the structural dynamics of water.[4,37,48,49]

We investigated the mechanism of slowing down the structural changes in water at ambient pressure by analyzing jump motions obtained from molecular dynamics (MD) simulations.[4,50] We found that the jump dynamics deviate from Poisson statistics with decreasing temperature due to dynamic disorder, in which slow motions compete with jump motions.[51-57] Previous studies have reported that molecules in the HDL-like state (i.e., molecules with an extra molecule in the first hydration shell) facilitate structural changes in water.[37, 48, 49] Therefore, one might expect this slowdown to be attributed to the fifth-nearest neighbor molecules of the jumping molecules. However, the present analysis revealed that the slow displacements of the fourth-nearest neighbor molecules of the jumping molecules play a pivotal role in slowing down the jump motions in supercooled water. We also found that the displacements of the fourth-neighbor molecules occur in the environment created by the fluctuations of molecules outside the first hydration shell. In addition, we found that with decreasing temperature, the molecular motions in the jumping process become increasingly cooperative. Consequently, the number of slow variables influencing the jump dynamics increases, leading to slower and more intermittent jump dynamics in a higher-dimensional space.

The organization of this paper is as follows. The theoretical and computational details are described in Sec. II. The results are presented and discussed in Sec. III. Finally, the conclusions are summarized in Sec. IV.

## II. THEORETICAL AND COMPUTATIONAL DETAILS
### A. MD simulations

In this study, we used the MD trajectory data generated by our program in previous research.[4, 50] Detailed information on the MD simulations can be found in the previous reports and is summarized here. First, we performed MD simulations of liquid water at 1 atm with 1,000 water molecules under constant temperature and constant pressure (NPT) conditions to determine the densities at 300, 250, 230, 215, 205, and 200 K. The system size used in this study is not particularly large. However, as shown in Fig. 3, it is sufficiently large compared to the correlation length, which increases with decreasing temperature. Therefore, it is considered that the current system size does not significantly affect the conclusions drawn in this study. The TIP4P/2005 model potential[47] was used for the water molecules, and the periodic boundary condition was imposed. Long-range electrostatic interactions were calculated using the Ewald summation. The Nóse−Hoover thermostat and Berendsen barostat were used for the NPT simulations. The obtained densities[50] were in good agreement with those reported by Pi et al.[58] Subsequently, we performed extensive MD simulations at the previously determined densities under constant volume and constant energy (NVE) conditions to reduce fluctuations induced by the thermostat and barostat. A time step of 1 fs was used for the time evolution of the translation and rotation motions, and the trajectory data were recorded at 10 fs intervals. The calculated average temperatures were 300, 250, 230, 215, 205, and 197 K. A number of properties were analyzed at these temperatures, including heat capacity, isothermal compressibility, intermediate scattering function, and relaxation time of structural changes,[4, 50] all of which were consistent with the results of previous studies.[35, 47, 58] The trajectory lengths used in the present analysis were 10, 20, 108, 120, 135, and 540 ns for 300, 250, 230, 215, 205, and 197 K, respectively. The dynamics of structural changes significantly slow down below ~200 K, making the present analyses, which use long datasets with short time intervals, practically challenging. Therefore, this study focused on performing comprehensive analyses of the jump dynamics in the temperature range between 300 and 197 K. To investigate the ensemble dependence of the present results, we performed the MD simulations of liquid water under the NVT ensemble conditions using the Nosé-Hoover thermostat. As a result, we found no significant difference in the statistics of jump dynamics between the NVT and NVE ensembles (see Sec. III. B).

All of the above water densities depend on temperature and are less than 1 g/cm$^3$. To investigate the change in water dynamics due to an increase in density, we also used the MD trajectory data of liquid water under the NVE conditions at a constant density of 1 g/cm$^3$ at 300, 250, 230, 215, 205, and 190 K (Ref. [50]). The trajectory lengths at 300, 250, 230, 215, 205, and 190 K were 10, 18, 75, 105, 120, and 150 ns, respectively, and the trajectory data were saved at 10-fs intervals.

### B. Identifying jump motion

Several methods have been proposed to study jump motions.[10-12, 14, 16-18] In the present study, we used the method proposed by Candelier et al.,[10] in which jumps of molecule $i$ are analyzed using a hop function, $h_i(t)$,

$$h_i(t) = \sqrt{\langle(\mathbf{r}_i(t) - \langle\mathbf{r}_i(t)\rangle_B)^2\rangle_A \langle(\mathbf{r}_i(t) - \langle\mathbf{r}_i(t)\rangle_A)^2\rangle_B} \,. \qquad (1)$$

Here $\mathbf{r}_i$ represents the geometrical center of water molecule $i$, $\mathbf{r}_i = \frac{1}{3}\sum_{a=1}^{3} \mathbf{r}_{i,a}$, to take into account the translational and orientational displacements. $\langle X \rangle_A$ and $\langle X \rangle_B$ are the averages of $X$ in the time windows A and B, which are defined as $\langle X(t)\rangle_A = \frac{1}{\Delta t/2}\int_{t-\Delta t/2}^{t} X(\tau)\, d\tau$ and $\langle X(t)\rangle_B = \frac{1}{\Delta t/2}\int_{t}^{t+\Delta t/2} X(\tau)\, d\tau$, respectively. $h_i(t)$ is the square root of the product of the average squared distances from the mean position in the different time windows. This function shows a rapid increase when the molecule undergoes a position change and a rapid decrease at the end of the jump [Fig. 1(b)]. To determine $\Delta t$ for time windows, the mean square displacement (MSD), $\langle\delta^2 r(t)\rangle$, has been employed.[10-14] In this study, we used the time of minimum diffusivity, i.e., the time when $d\log\langle\delta^2 r(t)\rangle/d\log t$ is minimized (Fig. S1).[13, 14, 18] At 205 and 197 K, the MSD shows oscillations in the time range from ~1 to ~30 ps, and $d\log\langle\delta^2 r(t)\rangle/d\log t$ shows the global minimum at ~6 ps for those two temperatures. However, when $\Delta t$ exceeds the ballistic timescale (~0.2 ps), a smaller $\Delta t$ provides greater temporal resolution for distinguishing between localized and jump motions.[12] Therefore, for these temperatures, $\Delta t$ was determined from the time of minimum diffusivity at which these oscillations were assumed to be absent. As shown later, there is no significant $\Delta t$ dependence in the present results as long as $\Delta t$ falls within the flat range of the MSD.[12] The $\Delta t$ used in the present study is summarized in Table SI.

To distinguish between the localized vibrational and jump motions (i.e., the cage and jump states), we analyzed the previously used cumulative probability of the hop function, which represents the fraction of displacements greater than a certain value of $h$.[12, 22] As shown in Figs. S2(a)−S2(c), the cumulative probability of the hop function rapidly decreases up to $h \sim 2.5$ Å$^2$, followed by a slower exponential tail associated with less frequent but larger jump motions. Consequently, we determined the hopping threshold between the localized and jump motions, $h^*$, to be 2.5 Å$^2$. In the present study, the temperature dependence of $h^*$ in water is almost negligible. In addition, we confirmed that the value of $h^*$ remains at 2.5 Å$^2$ in the $h$ histograms calculated for $\Delta t = 6$ ps at 205 and 197 K [Fig. S2(d)]. These results show that $h^*$ remains nearly constant as long as $\Delta t$ is neither too small (i.e., the ballistic timescale) nor too large (i.e., the structural relaxation timescale).[12]

## C. Participation fraction

We performed normal mode analysis by calculating the mass-weighted second derivative of the potential energy and obtained normal modes and their corresponding frequencies. The participation fraction, $\phi_{i,m}$, of a jumping molecule $i$ in normal mode $m$, whose eigenvector and eigenvalue are $\boldsymbol{e}_m$ and $\omega_m$, respectively, is given by[9]

$$\phi_{i,m} = \sum_\alpha |e_{i\alpha,m}|^2 , \qquad (2)$$

where the subscript $\alpha$ represents the degree of freedom, i.e., the translational and rotational components. In liquid water, the intermolecular translational and librational motions are found below ~400 cm$^{-1}$ and above ~400 cm$^{-1}$, respectively.[59] The participation fraction of molecule $i$ at frequency $\omega$ is defined as

$$w_i(\omega) = \langle \sum_m \phi_{i,m} \delta(\omega - \omega_m) \rangle , \qquad (3)$$

where the angular brackets denote the ensemble average. We obtained the participation fraction and the density of states by averaging over 1,000 configurations.

## D. Randomness parameter and survival probability

The randomness parameter, $R$, is defined as[60, 61]

$$R = \frac{\langle t^2 \rangle - \langle t \rangle^2}{\langle t \rangle^2}, \qquad (4)$$

where $\langle t^n \rangle = \int t^n \psi(t)\, dt$. In Eq. (4), $\psi(t)$ represents the residence time distribution of the cage state, i.e., a state with $h < h^*$. When a stochastic process follows Poisson statistics, $\psi(t)$ can be described by an exponential function [i.e., $\psi(t) = ke^{-kt}$], and $R$ equals unity.

Once we have $\psi(t)$, we can obtain the residence probability, $C_R(t)$, and the survival probability, $C_S(t)$,

$$C_R(t) = \int_t^\infty \psi(\tau)\, d\tau, \qquad (5)$$

$$C_S(t) = \frac{1}{\langle t \rangle} \int_t^\infty C_R(\tau)\, d\tau. \qquad (6)$$

When a Poisson process is assumed [i.e., $\psi(t) = ke^{-kt}$], both the residence and survival probabilities are identical, i.e., $C_R(t) = C_S(t) = e^{-kt}$.

The formal rate equation for the survival probability, $C_S(t)$, is given by[51]

$$\frac{dC_S(t)}{dt} = -k(t) C_S(t), \qquad (7)$$

because the rate, $k(t)$, generally fluctuates due to environmental effects. Its formal solution is expressed as

$$C_S(t) = \left\langle \exp\left(-\int_0^t k(\tau) d\tau\right) \right\rangle. \qquad (8)$$

We can derive two limits from the above formal solution. One is the fast fluctuation limit, where $k(t)$ fluctuates very quickly. In this limit, the survival probability is described by the exponential function[51]

$$C_{fast}(t) = \exp(-k_{fast} t), \qquad (9)$$

with an average rate $k_{fast}$. Here, we assumed that $(N_J + 1)$ jumps are observed during time $T$ discretized into $N_T$ steps with a time interval of $\Delta t$ ($T = N_T \Delta t$). Without loss of generality, we also assumed that the first jump occurs at the time of origin for simplicity. Therefore, the average rate $k_{fast}$ is given by $N_J/(N_T \Delta t)$, corresponding to the result of the flux-over-population method.[56, 57]

The second limit is the slow fluctuation limit, where $k(t)$ changes very slowly. In this limit, the survival probability is given by Eq. (10), represented as a multi-exponential function weighted by the distribution of $k$,[51]

$$C_{slow}(t) = \langle \exp(-kt) \rangle_k . \tag{10}$$

At the molecular level, the fluctuation of $k(t)$ is caused by slow variables other than the reaction coordinate, such as the position of a neighboring molecule. In other words, different values of the slow variable (i.e., substates) give different $k(t)$ values. Here, we assume that a slow variable is composed of $N$ substates, each with different rates. When we denote $N_{Ti}$ as the number of steps in which substate $i$ is selected from the total number of steps $N_T$, and $N_{Ji}$ as the number of steps in which a jump ($h$ exceeds $h^*$) occurs in substate $i$, the relative population and rate of substate $i$ are $p_i = N_{Ti}/N_T$ and $k_i = N_{Ji}/(N_{Ti}\Delta T)$, respectively.[56, 57] Consequently, the survival probability at the slow fluctuation limit is expressed as

$$C_{slow}(t) = \sum_{i=1}^{N} p_i \exp(-k_i t) \tag{11}$$

$$= \sum_{i=1}^{N} \frac{N_{Ti}}{N_T} \exp\left(-\frac{N_{Ji}}{N_{Ti}\Delta T}t\right), \tag{12}$$

which is dependent on the choice of the slow variable. It is noted that, in the fast fluctuation limit, since the transitions between the substates are much faster than the reaction, the rate of each substate is $k_{fast}$. Therefore, the survival probability given by Eq. (9) is obtained.

To analyze the survival probability in the slow fluctuation limit, $C_{slow}^{(n)}(t)$, using $r_{OOn}$, which represents the distance between the oxygen (O)-atom of a jumping molecule and its $n$-th nearest O-atom, we divided the variable $r_{OOn}$ into small regions at 0.025 Å intervals (i.e., substates). For the analysis of $C_{slow}^{(\cos(\mathbf{d}_j \cdot \mathbf{d}_n))}(t)$ and $C_{slow}^{(|\cos(\mathbf{n}_j \cdot \mathbf{n}_n)|)}(t)$, we divided these variables into small regions at 0.01 intervals. Here, $\mathbf{d}_j$ and $\mathbf{n}_j$ represent the normalized dipole moment and the normalized vector connecting the two hydrogen (H)-atoms of molecule $j$, respectively. We then calculated the relative population and jump rate for these substates.

**E Kullback-Leibler divergence**

As shown above, $k_i$ depends on both $N_{Ji}$ and $N_{Ti}$. Therefore, the difference between the distributions at $h^*$ and equilibrium for substate $i$ causes fluctuations in $k$. Therefore, analysis of these distributions provides insight into the mechanism of dynamic disorder. We examined the change in the distribution $P(r_{OOn},h)$ toward the hopping threshold $h^*$ using the Kullback-Leibler (KL) divergence[62] defined as

$$D_n(h) = \int dr_{OOn} P(r_{OOn}, h) \log(P(r_{OOn}, h)/P(r_{OOn}, h^*)) . \tag{13}$$

We also analyzed the difference between the distributions $P^{eq}(r_{OOn})$ and $P(r_{OOn},h^*)$,

$$D_{eq,n} = \int dr_{OOn} P^{eq}(r_{OOn}) \log(P^{eq}(r_{OOn})/P(r_{OOn}, h^*)) . \tag{14}$$

$D_n(h)/D_{eq,n}$ represents the change in the distribution $P(r_{OOn},h)$ toward $P(r_{OOn},h^*)$ along the jump motion $h$, resembling a normalized time correlation function. Consequently, this analysis enables the elucidation of the differences in the rate of change of the distribution dependent on $r_{OOn}$.

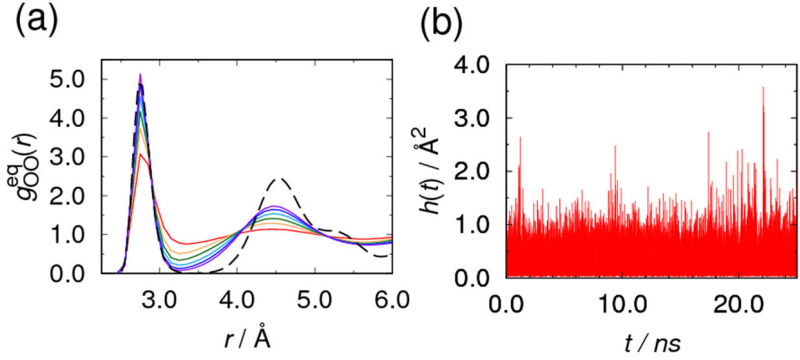

**FIG. 1**. (a) Temperature dependence of the RDF between the O-atoms of water molecules in liquid water. (b) Time series of the hop function of a molecule in liquid water at 197 K. In (a), the red, orange, green, light blue, blue, violet, and black-dashed curves represent the RDFs at 300, 250, 230, 215, 205, 197 K, and ice Ih at 250 K, respectively.

## III. RESULTS AND DISCUSSION

### A. Changes in structural properties during jump motions

We analyzed the displacements of individual water molecules using the hop function, $h$, [Eq. (1)].[10] Fig. 1(b) shows the time series of $h$ for a molecule in water at 197 K, a supercooled state below the Widom line (~220 K at ~ 1 atm), where the fractions of water molecules in the HDL-like and LDL-like states coincide.[4, 45] The $h$ values show long-lasting small fluctuations and sudden large jumps, corresponding to oscillatory motions within a cage and jumps between cages, respectively. As described in Sec. II. B, we distinguished between the cage and jump states at the point where the shape of the cumulative probability of $h$ changes, referred to as the hopping threshold $h^*$ (Fig. S2).[12]

Using $h$ as an order parameter for the jump motions, we obtained the free energy profile along $h$, $\Delta F(h)$, and found an increase in the free energy barrier for the jump motion from 2.8 kcal/mol [$\Delta F(h^*)/k_B T = 4.7$] at 300 K to 4.4 kcal/mol [$\Delta F(h^*)/k_B T = 11.2$] at 197 K [Fig. 2(a)].

We then analyzed the $h$ dependence of several static properties. Fig. 2(b) shows the RDF between the O-atoms of a jumping molecule with different $h$ values and the surrounding O-atoms, $g_{OO}(r,h)$, at 197 K. As $h$ increases, the first and second maxima decrease, while the first minimum increases (also see Figs. S3−S5). We also examined the $h$ dependence of CN, which is defined as the number of molecules within the first minimum of $g_{OO}(r,h)$. MD simulations have shown that the water density increases as the temperature rises from ~190 to ~270 K.[4, 27, 35] Therefore, it might be expected that higher values of $h$ would correspond to higher CN in this temperature range. However, the opposite has been observed: CN decreases with increasing $h$ [Figs. 2(c), 2(d), and S6]. Furthermore, we examined the $h$ dependence of the tetrahedral order parameter, $Q_i$, which indicates the extent to which the local structure of molecule $i$ resembles a tetrahedral arrangement. The $Q_i$ is defined as[63]

$$Q_i = 1 - \frac{3}{8}\sum_{j=1}^{3}\sum_{k=j+1}^{4}\left(\cos\theta_{jik} + \frac{1}{3}\right)^2, \qquad (15)$$

where the indices $j$ and $k$ represent two of the four nearest neighbor water molecules. As shown in Figs. 2(e) and S7, the peak of the $Q$ distribution is found at $Q > 0.8$ at equilibrium. However, as $h$ increases, the peak shifts to smaller $Q$ values and reaches ~0.5 at $h^*$, characterized by three- and two-coordinated water molecules.

The decrease in CN with increasing $h$ is supported by the participation fraction of jumping molecules in intermolecular vibrational normal modes (see Sec. II. C).[9, 64, 65] An increase in the density of states (DOS) between 300 and 400 cm$^{-1}$ for translational motion and a decrease in the DOS above 450 cm$^{-1}$ for librational motion are observed with increasing $h$ [Figs. 2(f) and S8] due to an increase in the fraction of three- and two-coordinated water molecules and a decrease in the fraction of four-coordinated water molecules (Fig. S9).[4] The present finding is also consistent with the results from machine learning techniques for other systems, which show that *soft* particles with fewer neighbors are more likely to undergo structural rearrangements than *hard* ones.[20, 21, 23]

We investigated the correlation of displacements by analyzing the average $h$ of molecules in each $n$-th hydration shell, $\bar{h}_n$, ($1 \leq n \leq 4$), of a jumping molecule [Figs. 3(a)−3(f)]. At all temperatures studied here, the displacement increased with proximity to the jumping molecule. In addition, the ratio of $\bar{h}_n$ to $h_{eq}$, the average $h$ of the system, increased with decreasing temperature [Figs. 3(g)−3(j)], showing a growth of the correlation length, reminiscent of the Adam-Gibbs theory.[66]

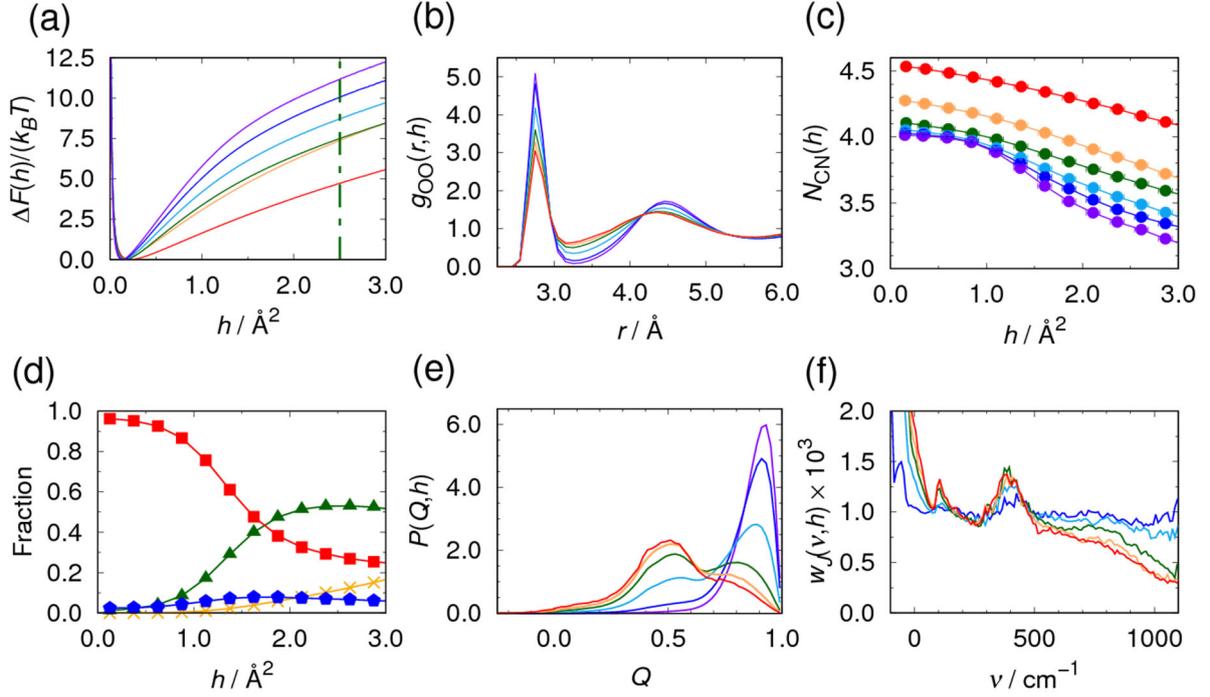

**FIG. 2**. (a) Temperature dependence of the free energy profile along $h$. (b) RDF between the O-atom of a jumping molecule with different $h$ values and its surrounding O-atoms in liquid water at 197 K. (c) Temperature dependence of the average CN for a jumping molecule with $h$. (d) Fractions of jumping molecules with 2 (orange)-, 3 (green)-, 4 (red)-, and 5 (blue)-coordination in liquid water at 197 K. The tetrahedral order parameter (e) and participation fraction (f) of a jumping molecule with different $h$ values in liquid water at 197 K. In (a), the vertical green dashed-dotted line represents the hopping threshold. In (a) and (c), the red, orange, green, light blue, blue, and violet curves represent the results for water at 300, 250, 230, 215, 205, and 197 K, respectively. In (b), (e), and (f), the curves are color-coded to represent results for a jumping molecule with specific $h$ value ranges: violet (0.25-0.50), blue (0.75-1.00), light blue (1.25-1.50), green (1.75-2.00), orange (2.25-2.50), and red (2.75-3.00) in (b) and (e), and blue (0.50-0.75), light blue (1.00-1.25), green (1.50-1.75), orange (2.00-2.25), and red (2.50-2.75) in (f).

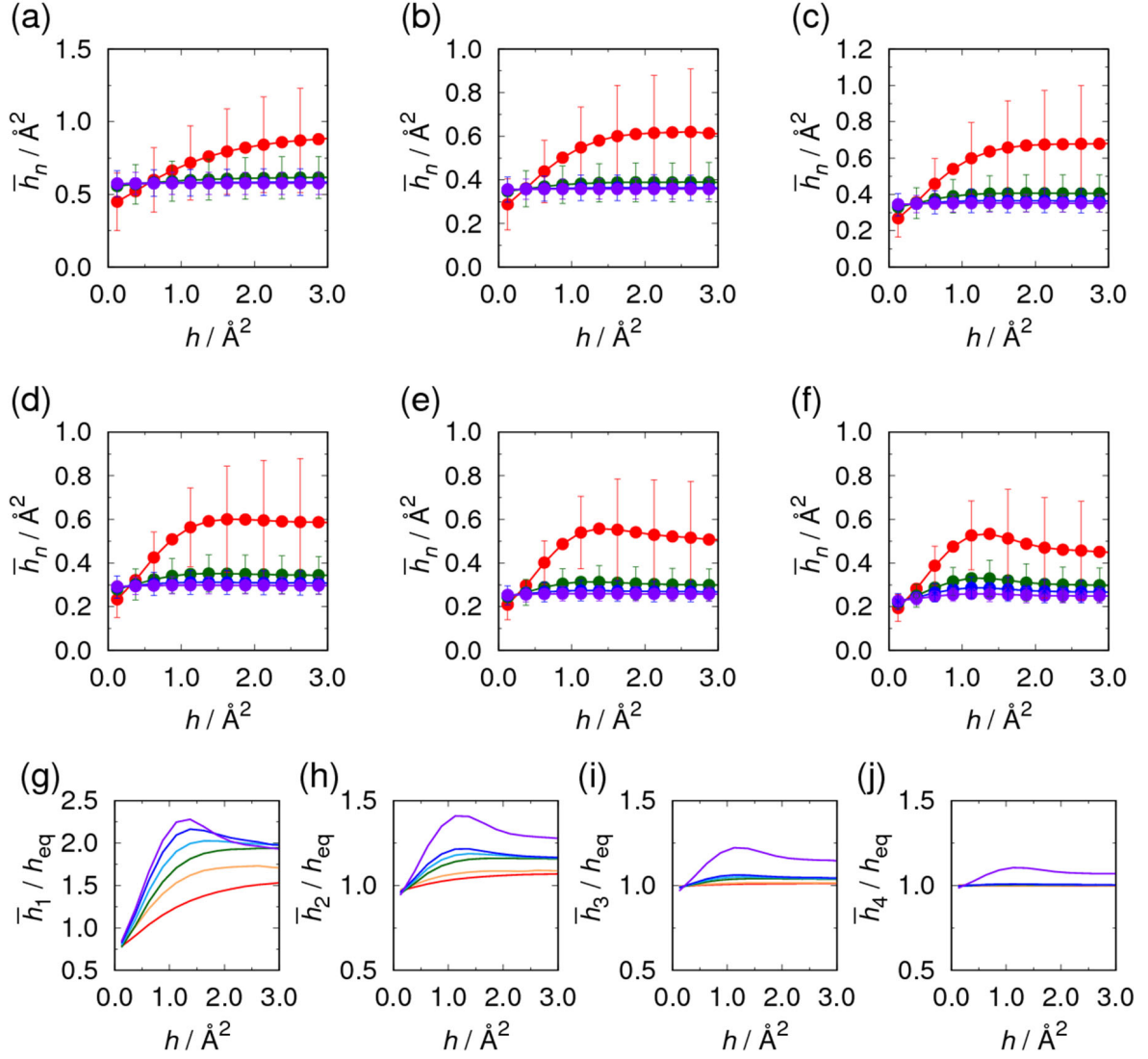

**FIG. 3**. Average value of $h$ for molecules in the first (red), second (green), third (blue), and fourth (violet) hydration shells of a jumping molecule with $h$ in liquid water at 300 (a), 250 (b), 230 (c), 215 (d), 205 (e), and 197 K (f). The ratio of the average of $h$ for molecules in the first (g), second (h), third (i), and fourth (j) hydration shell to the average $h$ of all molecules in liquid water, $h_{eq}$. In (a)−(f), error bars are shown in every two data for clarity. In (g)−(j), the red, orange, green, light blue, blue, and violet curves represent the results for liquid water at 300, 250, 230, 215, 205, and 197 K, respectively.

## B. Transition of the nature of jump motions

We next examined the residence time of the cage state, $\psi(t)$ [Fig. 4(a)]. While the residence time of the jump state is short and nearly temperature independent [Fig. 4(b)],[16] $\psi(t)$ is significantly long and temperature dependent. In addition to its increasing timescale with decreasing temperature, the behavior of $\psi(t)$ also changes. At $T \geq 250$ K, $\psi(t)$ follows an exponential function, while at lower temperatures, it transitions to a non-exponential function with a long tail, exhibiting non-Poisson statistical characteristics. To quantify the change in the nature of the jump dynamics, we analyzed the randomness parameter, $R$ [Eq. (4)].[60, 61] As shown by the red dots in Fig. 4(c), $R$ remains close to unity for $T \geq 250$ K, then gradually increases below ~250 K, surges at ~220 K, and reaches ~13 at ~200 K.

Fig. 4(c) shows that the temperature at which $R$ abruptly increases coincides with the Widom line, where the fractions of water molecules in the HDL-like and LDL-like states coincide. It has been shown recently that at temperatures below the Widom line, the clusters of molecules in the HDL-like state are fragmented and isolated within a percolating network of the LDL-like state, thereby slowing the structural dynamics.[4] Additionally, it has been reported that the LDL-like state is destabilized in supercooled water with a density of 1 g/cm$^3$.[50] Therefore, we investigated the jump dynamics in liquid water at this high density. Consequently, we found a shorter tail in $\psi(t)$ and a decrease in $R$ [Figs. 4(c) and 4(d)] due to the destabilization of the LDL-like state. This destabilization, caused by an increase in density, lowers the crossover temperature between the HDL-like and LDL-like states [Fig. 4(e)]. The results indicate that the slow, intermittent non-Poisson water dynamics are induced by the prolonged trapping of jumping molecules within a network of LDL-like states stabilized at low temperatures and low densities.

The shift in the fractions of water molecules in the HDL-like and LDL-like states leads to a change in the entropy of water. As the temperature decreases from room temperature, the entropy of liquid water sharply declines near the Widom line, due to the shift in the fraction of water molecules from the HDL-like state with higher entropy to the LDL-like state with lower entropy [red curves in Fig. 4(e)]. Additionally, it is known that the configurational entropy, $S_C$, and excess entropy, $S_{ex}$, of water rapidly decrease as the temperature lowers.[59, 67-75] Furthermore, it has been demonstrated that the decreases in $S_C$ and $S_{ex}$ are correlated with a reduction in the diffusion coefficient, using the Adam-Gibbs relation[66] and the Rosenfeld scaling[76].[70-75] Therefore, the present study reveals that the slow and intermittent non-Poisson dynamics underlie the decrease in the diffusion coefficient of water at low temperatures.

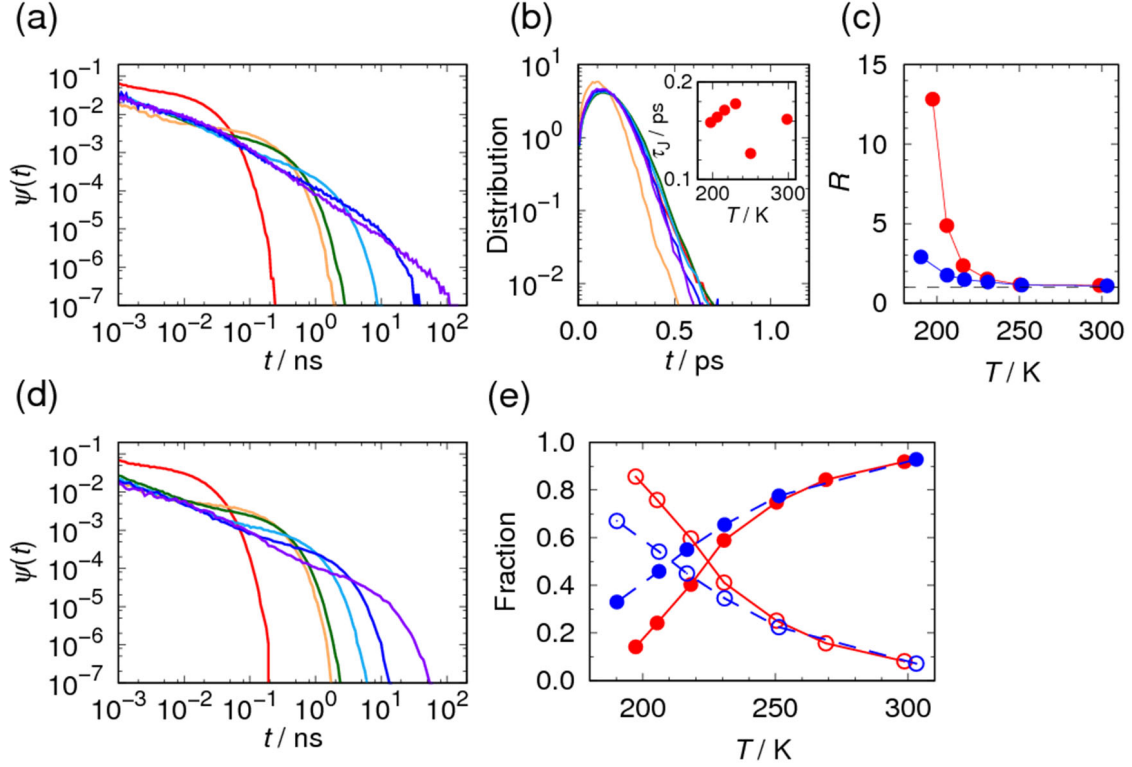

**FIG. 4.** Temperature dependence of the residence time distribution for the cage state, $\psi(t)$, (a), and the jump state (b). The inset of (b) shows the average residence time for the jump state. (c) Temperature dependence of the randomness parameter of liquid water. (d) Temperature dependence of $\psi(t)$ in liquid water with a density of 1 g/cm$^3$. (e) Temperature dependence of the fraction of the water molecules in the HDL-like (filled circle) and LDL-like states (hollow circle). In (a) and (b), the red, orange, green, light blue, blue, and violet curves represent $\psi(t)$ in liquid water at 300, 250, 230, 215, 205, and 197 K, respectively. In (d), the red, orange, green, light blue, blue, and violet curves represent $\psi(t)$ in liquid water at 300, 250, 230, 215, 205, and 190 K, respectively. $\psi(t)$ in (a) and the red points in (c) and (e) represent the results obtained from the MD simulations of liquid water at the density determined under constant-pressure conditions, whereas $\psi(t)$ in (d) and the blue points in (c) and (e) represent the results obtained from the MD simulations of liquid water with a density of 1 g/cm$^3$. The definitions for the LDL-like and HDL-like states in Ref. 4 were used in this study.

We then examined the jump dynamics using the survival probability of the cage state, $C_S(t)$, and its fast fluctuation limit, $C_{fast}(t)$ (Fig. 5). Here, $C_S(t)$ represents the probability that a molecule will not undergo a jump within time $t$ [see Sec. II. D and Eq. (6)]. We analyzed the survival probability calculated from the MD simulations with a stretched exponential function,

$$C_S(t) = \exp\left[-(kt)^\beta\right]. \qquad (16)$$

The rate, $k$, the exponent, $\beta$, and the rate in the fast fluctuation limit, $k_{fast}$, are summarized in Table I. $\beta$ represents the heterogeneity of the rate. At $T \geq 250$ K, $\beta$ is ~1, indicating that $k(t)$ is well approximated by $k_{fast}$. Consequently, $C_S(t)$ closely follows an exponential function and is closely approximated by $C_{fast}(t)$ [Figs. 5(a) and 5(b)]. These results indicate that the jump dynamics follow Poisson statistics ($R \sim 1$) and that variables other than $h$ relax significantly faster than $h$. In contrast, at temperatures below 250 K, $\beta$ gradually becomes smaller than 1, $k(t)$ fluctuates slowly, and the approximation of $k(t) \sim k_{fast}$ is no longer valid. As a result, $C_S(t)$ exhibits a long tail described by a stretched exponential function ($R > 1$), which is slower than $C_{fast}(t)$ expressed as $\exp(-k_{fast}t)$ [Figs. 5(c)−5(f)]. The results indicate that the barrier height for jump motions slowly fluctuates and that the jump dynamics can be described by a renewal process characterized by intermittent events [i.e., a non-exponential $\psi(t)$] arising from dynamic disorder.[51-54, 57]

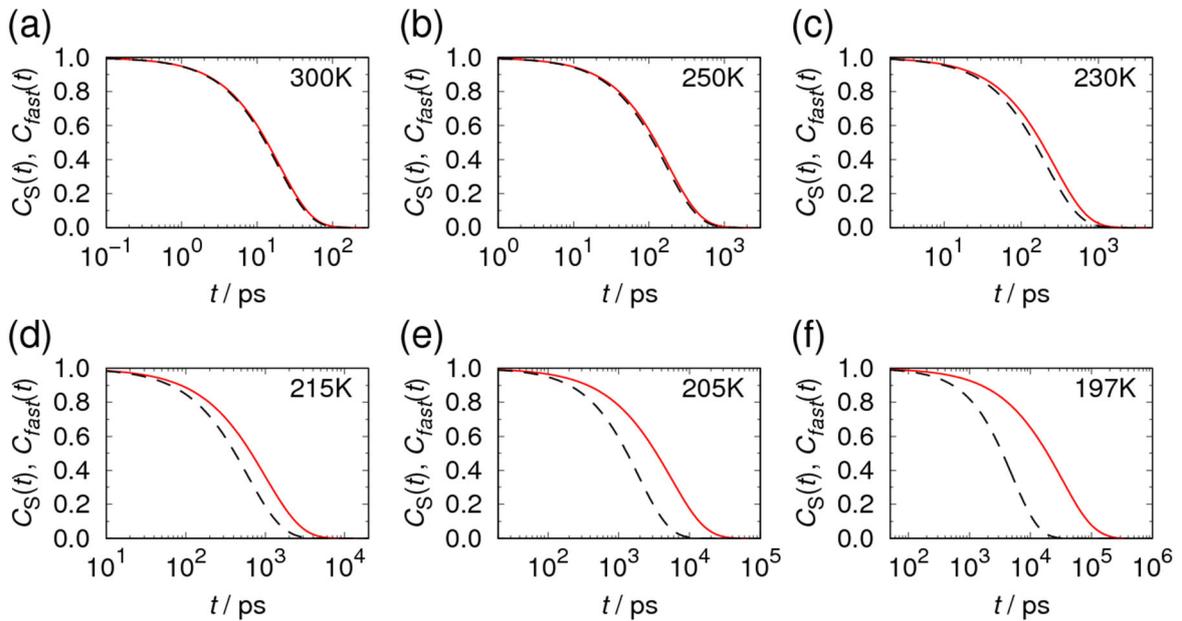

**FIG. 5.** Survival probability, $C_S(t)$ (red), and its fast fluctuation limit, $C_{fast}(t)$ (dashed black), of the cage state at 300 (a), 250 (b), 230 (c), 215 (d), 205 (e), and 197 K (f).

**Table I.** Rate, $k$, and exponent, $\beta$, fitted by a stretched exponential function, $\exp[-(kt)^\beta]$, to the survival probability of the cage state in liquid water at 300, 250, 230, 215, 205, and 197 K, along with the rate, $k_{fast}$, in the fast fluctuation limit represented by $\exp(-k_{fast}t)$.

| $T$ (K) | 300 | 250 | 230 | 215 | 205 | 197 |
|---|---|---|---|---|---|---|
| $k$ (ps$^{-1}$) | $5.08 \cdot 10^{-2}$ | $5.55 \cdot 10^{-3}$ | $3.78 \cdot 10^{-3}$ | $1.01 \cdot 10^{-3}$ | $1.92 \cdot 10^{-4}$ | $3.31 \cdot 10^{-5}$ |
| $\beta$ | 0.993 | 0.996 | 0.976 | 0.931 | 0.851 | 0.763 |
| $k_{fast}$ (ps$^{-1}$) | $5.35 \cdot 10^{-2}$ | $5.99 \cdot 10^{-3}$ | $4.72 \cdot 10^{-3}$ | $1.66 \cdot 10^{-3}$ | $5.25 \cdot 10^{-4}$ | $2.00 \cdot 10^{-4}$ |

We also investigated the temperature dependence of the survival time of $C_S(t)$. Here, the survival time is defined as the time at which $C_S(t) = 1/e$. Fig. 6 shows all the survival times obtained in this study along with the results of fitting the survival times using the Vogel-Fulcher-Tammann (VFT) equation for two temperature ranges, above 215 K and below 230 K. The VFT equation is expressed by

$$\tau(T) \sim \exp[DT_0/(T - T_0)] . \tag{17}$$

Here, $D$ and $T_0$ denote the fragility index and the VFT temperature, respectively. The values of these parameters for the two temperature ranges are presented in Table II. A larger fragility index indicates more Arrhenius-like behavior. Therefore, we found a transition from a marked non-Arrhenius dependence to a weak non-Arrhenius behavior at ~220 K, i.e., the Widom line at $P = 1$ atm. A similar transition in the fragility of water near the Widom line has been reported in MD simulations of liquid water employing various model potentials.[4, 50, 77-80]

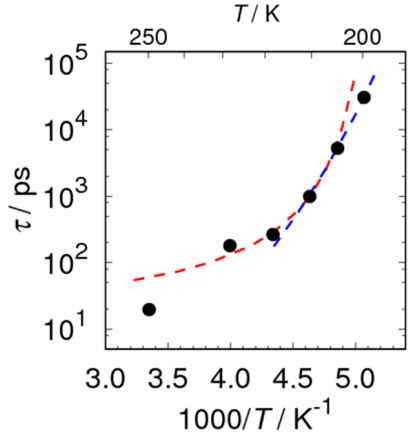

**FIG. 6.** Temperature dependence of the survival time for the cage state. The red and blue dashed curves represent the fits of the survival times for the two temperature ranges, above 215 K and below 230 K, by the VFT equation.

**Table II.** Fitting parameters of the VFT equation for the two temperature ranges, above 215 K and below 230 K.

|  | $D$ | $T_0$ (K) |
|---|---|---|
| 215 K $\leq T \leq$ 300 K | 0.62 | 185.6 |
| 197 K $\leq T \leq$ 230 K | 9.34 | 125.8 |

Before concluding this subsection, we discuss the influence of the chosen ensemble on the statistics of jump dynamics. Fig. S10 shows $\psi(t)$, $R$, and $C_S(t)$ for liquid water at temperatures above 200 K obtained from the MD simulations under the NVT ensemble conditions using the Nosé-Hoover thermostat. These results are in good agreement with those under the NVE ensemble conditions. Furthermore, it has been demonstrated that velocity scaling methods, particularly the Nosé-Hoover and stochastic rescaling thermostats, produce transport properties and thermodynamic distributions that are statistically indistinguishable from those under the NVE ensemble conditions.[81] Therefore, it is considered that the MD simulations conducted under the NVT ensemble conditions with an appropriate velocity scaling method yield results comparable to those obtained in this study.

## C. Microscopic origin of dynamic disorder in jump motions

As shown in Sec. III. A, the RDFs of jumping molecules change along $h$. Therefore, to elucidate the origin of the dynamic disorder, we analyzed the changes in the distance distributions of the molecules in the vicinity of the jumping molecules. First, we compared the distribution of the distance between the O-atom of the jumping molecule with $h$ and its $n$-th nearest O-atom, $P(r_{OOn}, h^*)$, to the distribution of the distance between the O-atom and its $n$-th nearest O-atom at equilibrium, $P^{eq}(r_{OOn})$, for $1 \leq n \leq 8$ [Figs. 7(a), 7(b), and S11]. We found that $P(r_{OOn}, h^*)$ shifts to longer distances compared to $P^{eq}(r_{OOn})$ for $n \leq 4$ but to shorter distances for $n \geq 5$. Notably, the largest difference between these two distributions is found at $n = 4$, where the average distance at $h^*$ exceeds the equilibrium distance by more than 0.4 Å at 197 K [Fig. 7(c)]. These shifts result in changes in $g_{OO}(r,h)$ and CN of the jumping molecule along $h$.

We then examined the change in $P(r_{OOn}, h)$ toward $P(r_{OOn}, h^*)$ using the KL divergence, $D_n(h)$ [see Sec. II. E and Eq. (13)]. Consistent with the above result, $D_4(h)$ is larger than other $D_n(h)$ values [Figs. 7(d) and S12(a)−S12(f)]. The rate of change from $P(r_{OOn}, h)$ to $P(r_{OOn}, h^*)$ was also examined using the scaled KL divergence, $D_n(h)/D_{eq,n}$. Here, $D_{eq,n}$ represents the KL divergence between $P(r_{OOn}, h^*)$ and $P^{eq}(r_{OOn})$ [see Eq. (14) and Figs. 7(e) and S12(g)−S12(l)]. We found that the $D_n(h)/D_{eq,n}$ decreases faster for $n \geq 5$ than for $n \leq 4$. This result indicates that the displacements of the inner four O-atoms occur in the environment created by the fluctuations of the outer molecules surrounding these four molecules in supercooled water. We also found a stronger temperature dependence of $D_n(h)/D_{eq,n}$ for $n \leq 4$ than for $n \geq 5$, which suggests that the displacements for $n \leq 4$ require a longer timescale at lower temperatures.

Conventional rate theories, such as transition state theory, assume that variables other than the reaction coordinate relax rapidly, resulting in a constant reaction rate independent of the variables and an exponential survival probability.[82] However, this assumption fails for supercooled water below 250 K due to dynamic disorder. Assuming that the fluctuations in $r_{OOn}$ ($1 \leq n \leq 8$) are as slow as the jump dynamics, we found that as the temperature decreases, the rate of jump dynamics becomes strongly dependent on the values of $r_{OO4}$ [the solid green line in Fig. 7(a)]; for example, the ratio of the rate at the peak of $P(r_{OO4},h^*)$ to that at the peak of $P^{eq}(r_{OO4})$ is ~320, which is significantly larger than the ratios of other $n$ values [Figs. 7(a), 7(b), and S13 and Table III]. Therefore, the displacement of the fourth-nearest O-atom of the jumping molecule, $r_{OO4}$, is the most relevant slow variable for the dynamic disorder in the jump dynamics.

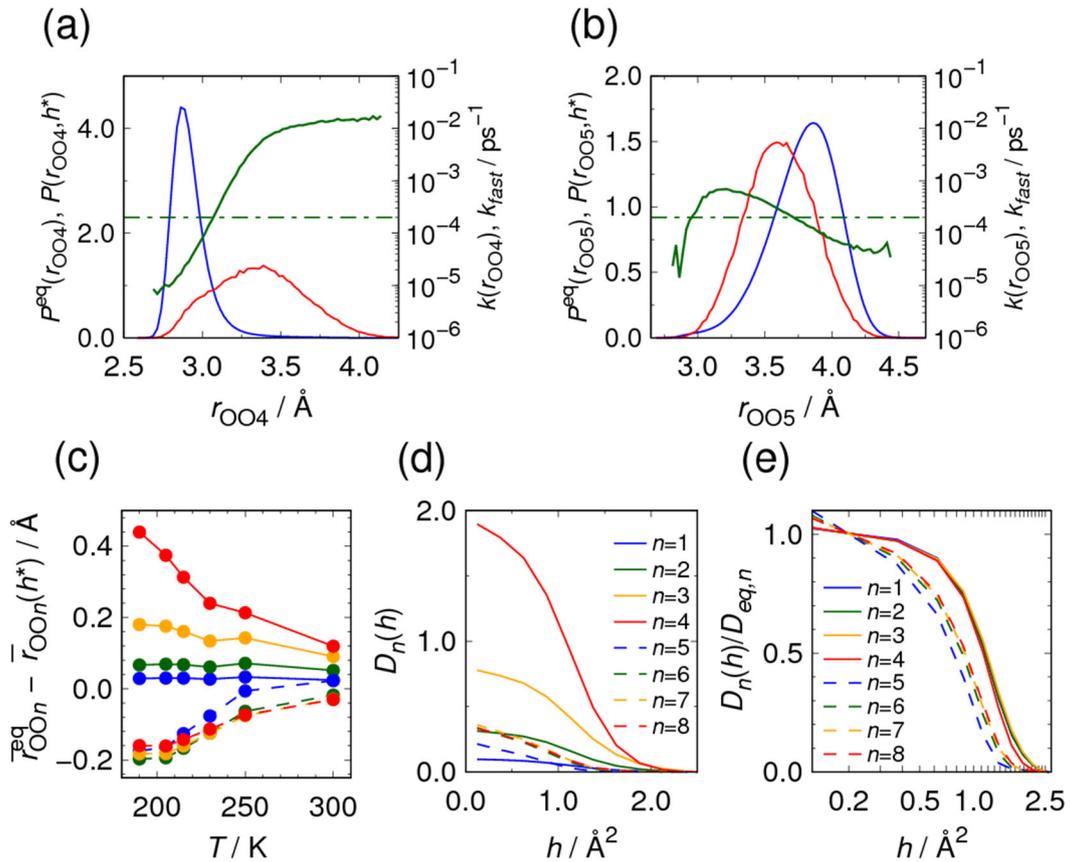

**FIG. 7**. (a) Distributions of the distances between the O-atom and its fourth-nearest O-atom in equilibrium (blue) and between the O-atom of a jumping molecule with $h^*$ and its fourth-nearest O-atom (red) in liquid water at 197 K. The green curve and green dashed-dotted line represent $k(r_{OO4})$ and $k_{fast}$, respectively. (b) Distributions of the distance between the O-atom and its fifth-nearest O-atom in equilibrium (blue) and between the O-atom of a jumping molecule with $h^*$ and its fifth-nearest O-atom (red) in liquid water at 197 K. The green curve

and green dashed-dotted line represent $k(r_{OO5})$ and $k_{fast}$, respectively. (c) Temperature dependence of the difference between the average distances at equilibrium and $h^*$ for the eight nearest neighbor molecules. KL divergence (d) and scaled KL divergence (e) for the eight nearest neighbor molecules in liquid water at 197 K. In (d) and (e), the solid (dashed) blue, green, orange, and red curves represent the results for $n$ = 1 (5), 2 (6), 3 (7), and 4 (8), respectively.

**Table III.** Ratios of the rate at the peak of $P(r_{OOn}, h^*)$ to that at the peak of $P^{eq}(r_{OOn})$.

| $T$ (K) | $n$ | | | | | | | |
| --- | --- | --- | --- | --- | --- | --- | --- | --- |
| | 1 | 2 | 3 | 4 | 5 | 6 | 7 | 8 |
| 300 | 1.04 | 1.20 | 1.43 | 1.97 | 1.02 | 1.00 | 1.05 | 1.06 |
| 250 | 1.09 | 1.38 | 1.96 | 5.33 | 1.00 | 1.18 | 1.27 | 1.16 |
| 230 | 1.04 | 1.35 | 1.75 | 7.38 | 1.21 | 1.60 | 1.67 | 1.51 |
| 215 | 1.09 | 1.42 | 2.29 | 23.68 | 1.76 | 2.49 | 2.09 | 2.29 |
| 205 | 1.13 | 1.51 | 2.40 | 100.00 | 2.59 | 3.08 | 2.72 | 2.85 |
| 197 | 1.12 | 1.53 | 2.45 | 321.86 | 2.34 | 2.83 | 3.13 | 2.53 |

We investigated the effect of fluctuations in $r_{OOn}$ on the jump dynamics by analyzing the slow fluctuation limit of the survival probability, $C_{slow}^{(n)}(t)$, considering $r_{OOn}$ as a slow variable that competes with $h$ [see Sec. II. D]. As mentioned earlier, for $T \geq 250$ K, $C_{fast}(t)$ closely approximates $C_S(t)$. However, $C_{slow}^{(4)}(t)$ decays more slowly than $C_S(t)$ [Figs. 8(a) and 8(b)]. This result indicates that $r_{OO4}$ remains a fast variable and that a one-dimensional reaction coordinate $h$ is sufficient to describe the jump dynamics. With decreasing temperature, $C_S(t)$ becomes slower and gradually approaches $C_{slow}^{(4)}(t)$ [Figs. 8(c)−8(e)], indicating that the jump dynamics cannot be described by $h$ alone and involve at least one more slow variable, $r_{OO4}$. However, the variable still fluctuates faster than the ideal slow limit. At 197 K, $C_{slow}^{(4)}(t)$ approximates $C_S(t)$ well, although it decays faster than $C_S(t)$ until ~3×10³ ps. Fig. 8(f) shows that $C_{slow}^{(4-5)}(t)$, where both $r_{OO4}$ and $r_{OO5}$ are considered slow variables, decays more slowly than $C_{slow}^{(4)}(t)$. Note that $C_{slow}^{(4-5)}(t)$ is slower than other $C_{slow}^{(4-n)}(t)$ ($6 \leq n \leq 8$) (Fig. 9), even though $C_{slow}^{(5)}(t)$ is faster than $C_{slow}^{(n)}(t)$ and $D_5(h)$ is smaller than $D_n(h)$. This is due to the stronger cooperativity between the displacements of the fourth and fifth molecules than between the displacements of the fourth and $n$-th molecules ($6 \leq n \leq 8$) (Fig. 10). The present result suggests that additional slow variables, such as $r_{OO5}$, are necessary for the jump dynamics at such low temperatures. As the temperature continues to decrease, the increased cooperativity of motions (i.e., an increase in the dimensionality of the jump dynamics) would further enhance slowing and intermittency, eventually culminating in a glass transition.

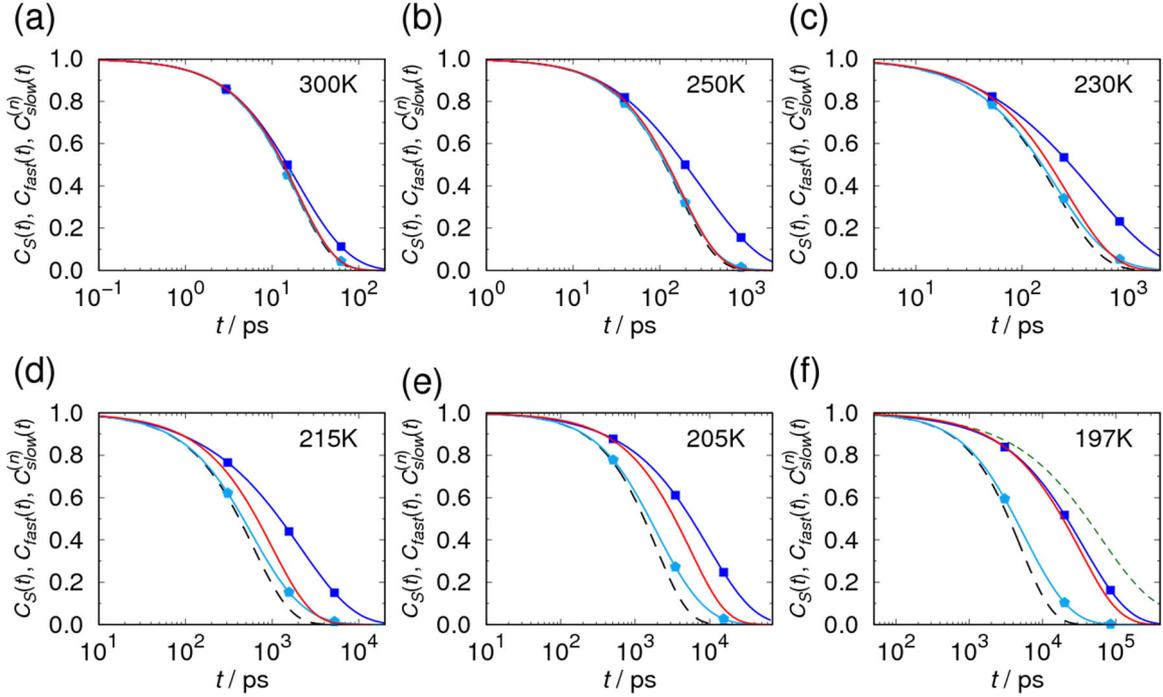

**FIG. 8.** Survival probability for the cage state, $C_S(t)$, its fast fluctuation limit, $C_{fast}(t)$, and the slow fluctuation limit, $C_{slow}^{(n)}(t)$, in liquid water at 300 (a), 250 (b), 230 (c), 215 (d), 205 (e), and 197 K (f). The red curve, dashed black curve, blue curve with squares, and light blue curve with pentagons represent $C_S(t)$, $C_{fast}(t)$, $C_{slow}^{(4)}(t)$, and $C_{slow}^{(5)}(t)$, respectively. In (f), the green dashed curve represents $C_{slow}^{(4\text{-}5)}(t)$.

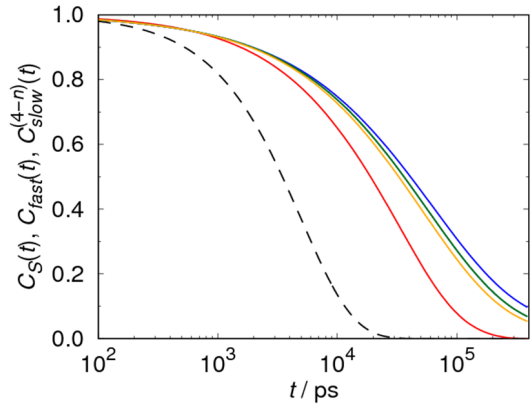

**Fig. 9.** Survival probability for the cage state, $C_S(t)$, and its fast and slow fluctuation limits in liquid water at 197 K. The solid red and dashed black curves represent $C_S(t)$ and $C_{fast}(t)$,

respectively. The blue, light blue, green, and orange curves represent the result for $n = 5, 6, 7$, and 8 in $C_{slow}^{(4-n)}(t)$, respectively. The light blue and green curves almost overlap.

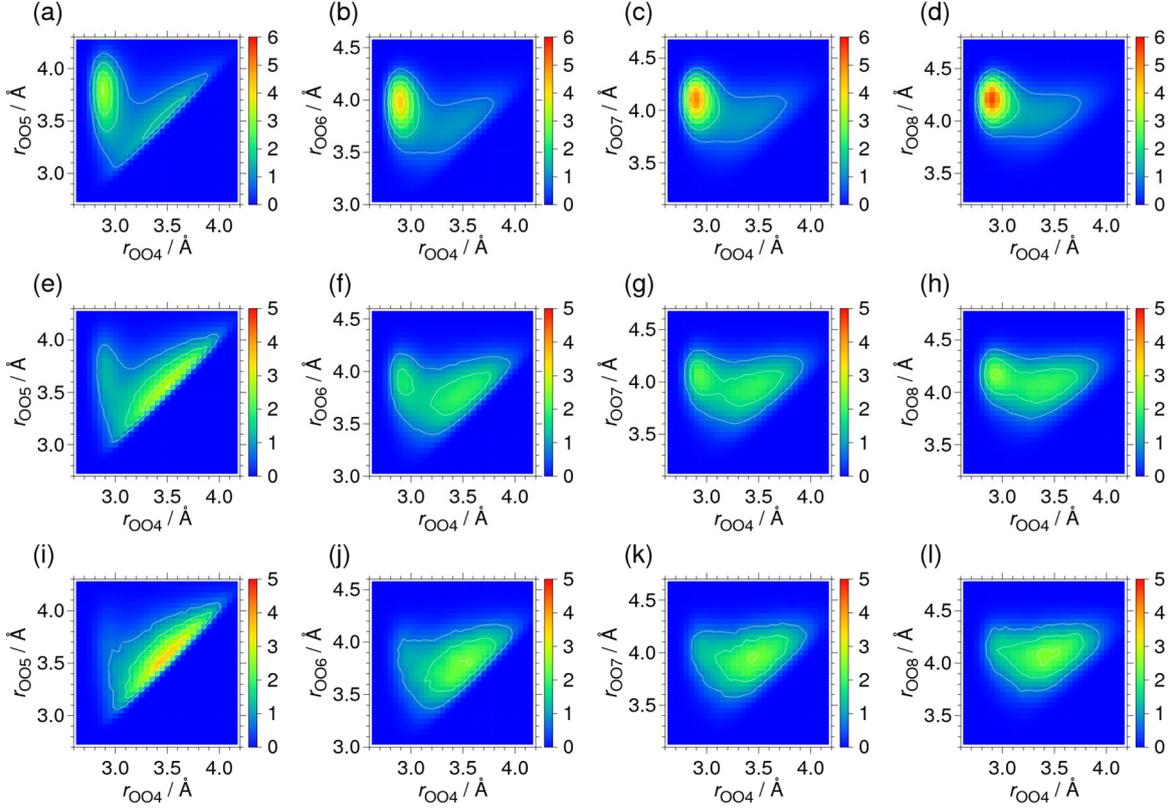

**Fig. 10.** Two-dimensional distributions of $r_{OO4}$ and $r_{OOn}$ ($5 \leq n \leq 8$) along $h$ in liquid water at 197 K. The range of the $h$ value is from 1.25 to 1.50 in (a)−(d), from 1.75 to 2.00 in (e)−(h), and from 2.25 to 2.50 in (i)−(l). All panels share the same color code, determined by the maximum value in (d).

We next examined the effect of the displacements involving H-atoms on dynamic disorder. First, we focused on $r_{OHn}$, $r_{HOn}$, and $r_{HHn}$ ($1 \leq n \leq 8$) (Figs. S14−S16). For example, $r_{OHn}$ represents the distance between the O-atom and its $n$-th nearest H-atom. We found noticeable changes in the distributions of $r_{OH2}$ and $r_{HO2}$, where $r_{OH2}$ represents the distance between the second-nearest H-atom and the O-atom of the jumping molecule, and $r_{HO2}$ represents the distance between the second-nearest O-atom and the H-atom of the jumping molecule. Changes in $r_{HHn}$ were found for $n = 7$ and 8. All four changes are related to the displacement of the fourth-nearest neighbor molecule of the jumping molecule. In addition, we examined the effects of two angles that depend on the position of the H-atoms: the angle between the dipole moments of the jumping and neighboring molecules and the angle between the vectors connecting the two H-atoms of these molecules. Differences between the equilibrium and $h^*$ distributions were found in the angles of the first four nearest neighbor molecules (Figs. S17 and S18). However, these angles have minimal influence on the dynamic disorder in the jump dynamics, as indicated by the similarity between the fast and slow fluctuation limits of the survival probability of these angles (Figs. 11 and S19). These results suggest that the displacements of the O-atoms play a primary role in the dynamic disorder of the jump dynamics.

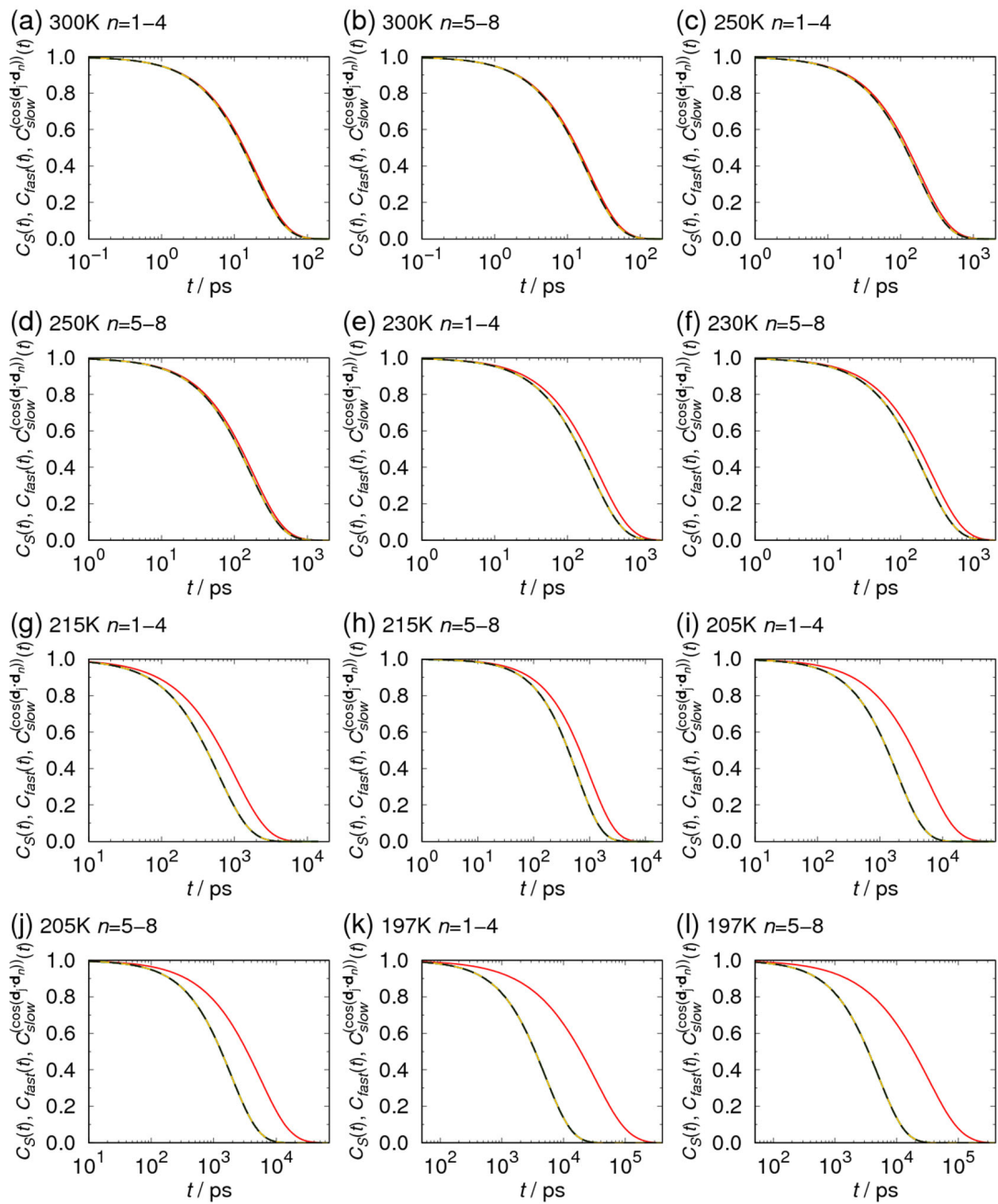

**Fig. 11.** Survival probability of the cage state, $C_S(t)$, in liquid water at 300 [(a) and (b)], 250 [(c) and (d)], 230 [(e) and (f)], 215 [(g) and (h)], 205 [(i) and (j)], and 197 K [(k) and (l)], along with their corresponding fast and slow fluctuation limits. The solid red and dashed black curves in all the panels represent $C_S(t)$ and $C_{fast}(t)$, respectively. In the slow fluctuation limit, the angle between the dipole moment of a jumping molecule with $h$ and that of its $n$th nearest neighbor molecule ($1 \leq n \leq 8$) is considered a slow variable. The blue, light blue, green, and orange curves in (a), (c), (e), (g), (i), and (k) represent the results for $n$ = 1, 2, 3, and 4, respectively, and those in (b), (d), (f), (h), (j), and (l) represent $n$ = 5, 6, 7, and 8. These curves overlap with $C_{fast}(t)$.

## IV. CONCLUSIONS

In summary, we investigated the microscopic mechanisms of the slowdown of the jump dynamics of supercooled water by utilizing the concept of dynamic disorder from reaction theory. By studying the survival probability and its fast and slow limits, we found that the displacements of the fourth-nearest O-atoms of the jumping molecules play a pivotal role in the dynamic disorder affecting the jump dynamics. It is noteworthy that the slow displacements of the fourth-nearest O-atoms occur within the environment created by the fluctuations of molecules located outside of the first hydration shell in supercooled water. Structural changes in the prearranged environment found here have also been observed in other systems, such as translational jumps of hydrophobic solutes in water and isomerization processes in a protein.[15, 83] At temperatures below 220 K, where the LDL-like state becomes dominant, the jump dynamics become slow and intermittent due to the trapping of the jumping molecules within an extended network of the LDL-like state. As the temperature continues to decrease, the jump dynamics become slower and more intermittent due to the increased cooperativity among molecular motions, i.e., the increased dimensionality of the jump dynamics. The present analysis thus provides insights into the molecular mechanisms underlying the slowing of the jump dynamics of liquid water, i.e., a significant decrease in the diffusion coefficient.

The analytical methods employed in this study (e.g., the survival probability and the randomness parameter) are not restricted to water. Slow motions cause slow fluctuations in the energy barriers, which, in turn, lead to fluctuations in the jump rate. Systems with different fragilities are expected to have different energy landscapes. Therefore, it is of interest to study the mechanisms of slowing down in systems with different fragilities. This method can elucidate the mechanism of slowing down in liquids in terms of the formation and migration of jump molecules (i.e., defects). Therefore, it would be intriguing to compare the results of this method with theories of glass slowing down and glass transition.[3, 19, 84-88] Additionally, non-Poisson dynamics are also observed in the conformational dynamics and reactions of proteins.[55-57, 89] Hence, this method is expected to elucidate the origin of complex dynamics in various systems, including biomolecular ones.

## SUPPLEMENTARY MATERIAL

See the supplementary material for additional details referenced in the text.

## ACKNOWLEDGMENTS

We acknowledge Prof. B. Bagchi, Prof. I. Ohmine, Dr. Z. Tang, Dr. R. S. Singh, and Dr. T. Inagaki for their valuable discussions. We are grateful to Dr. Tang for his guidance on efficient computation methods, which significantly contributed to the advancement of this study. The present study was supported by the Grant-in-Aid for Scientific Research (Grant Nos. JP21H04676 and JP23K17361) and DAICEL. The calculations were partially carried out using the supercomputers at the Research Center for Computational Science in Okazaki (Project Nos. 23-IMS-C196 and 24-IMS-C193).

## AUTHOR DECLARATIONS

### Conflict of Interest

The author has no conflicts to disclose.

### Author Contributions

**Shinji Saito**: Conceptualization; Data curation; Formal analysis; Investigation; Methodology; Project administration; Writing

## DATA AVAILABILITY

The data that support the findings of this study are available from the corresponding author upon reasonable request.